\documentclass[twocolumn,pre,superscriptaddress]{revtex4}

\usepackage{amsfonts}
\usepackage{amsmath,amssymb}
\usepackage[dvipdfm]{graphicx}

\begin{document}

\title{Analysis of instabilities and pattern formation in
time fractional reaction-diffusion systems}
\thanks{Preprint submitted to ASME}
\author{B.Y. Datsko}
\email{b_datsko@yahoo.com} \affiliation{Institute for Applied
Problems in Mechanics and Mathematics, National Academy of
Sciences of Ukraine, Naukova Street 3b, Lviv, Ukraine 79053}

\author{V.V. Gafiychuk}
\email{vagaf@yahoo.com} \affiliation{Institute for Applied
Problems in Mechanics and Mathematics, National Academy of
Sciences of Ukraine, Naukova Street 3b, Lviv, Ukraine 79053}

\date{\today}

\begin{abstract}
We analyzed conditions for Hopf and Turing instabilities to occur
in two-component fractional reaction-diffusion systems. We showed
that the eigenvalue spectrum and fractional derivative order
mainly determine the type of instability and the dynamics of the
system. The results of the linear stability analysis are confirmed
by computer simulation of the model with cubic nonlinearity for
activator variable and linear dependance for the inhibitor one. It
is shown that pattern formation conditions of instability and
transient dynamics are different than for a standard system. As a
result, more complicated pattern formation dynamics takes place in
fractional reaction-diffusion systems.
\end{abstract}

\maketitle
\affiliation{Physics Department, New York City College of Technology, CUNY, 300 Jay
Street, Brooklyn, NY 11201}
\affiliation{Institute for Applied Problems in Mechanics and Mathematics, National
Academy of Sciences of Ukraine, Naukova Street 3b, Lviv, Ukraine, 79053}

\section{Introduction}

 In reaction-diffusion systems a stable
equilibrium solution usually changes spontaneously with parameters
to limit cycle by Hopf bifurcation or stationary dissipative
structures by Turing bifurcation. As a result, we obtain nonlinear
dynamics leading to stationary or oscillatory structures. When
conditions of both instabilities arise, we can expect more complex
dynamics \cite{pr,ch,KO}. In case of a fractional
reaction-diffusion (FRD)
system dynamics can be much more complex \cite%
{hw1,lhw,hw2,gd,gd07,gd08,cam,siam,Nec}.

Last investigation showed that many complex heterogenous systems
are described by differential equations with fractional
derivatives to represent their anomalous behavior
\cite{zz,kl,mach,KST,Uch,ip}.\ Therefore, investigation of pattern
formation in FRD system has both theoretical and applied interest.

In this article the FRD system with cubic nonlinearity is studied
for the case when both instabilities take place. We have focused
on the dynamics of FRD model under conditions when sufficiently
complex patterns arise in the system dynamics. The results of
analytical treatment of the linearized model are validated by
computer simulations of nonlinear dynamics.

\section{MATHEMATICAL MODEL}

\label{sec:2}Let us consider the FRD system
\begin{equation}
\tau \;_{c}u_{t}^{^{\alpha }}=lu_{xx}+W(u,\mathcal{A}),  \label{1}
\end{equation}%
with two variables $u=(u_{1,}u_{2})^{T}$ on the $x\in
{(0,\mathcal{L})}$ subject to Neumann:
$u_{x}|_{x=0,\mathcal{L}_{x}}=0$ boundary conditions and with
certain initial conditions, $W=(W_{1},W_{2})^{T}$, $\quad
W_{1},W_{2}$ - smooth reaction kinetics functions, $\mathcal{A}$ -
real parameter, $\tau $
and $l$ are positive diagonal matrices $\tau =diag[{\tau _{i}}],$ $l=diag[{%
l_{i}^{2}}]>0$ .

Fractional derivatives $_{c}u_{t}^{^{\alpha }}$ on the left hand
side of the equations (\ref{1}), instead of the standard time
derivatives, are the Caputo fractional derivatives in time
\cite{ip,skm} of the order $0<\alpha <2 $ and are represented as
\begin{equation}
_{c}u_{t}^{^{\alpha }}=\frac{\partial _{c}^{\alpha }u(t)}{\partial
t^{\alpha
}}:=\frac{1}{\Gamma (m-\alpha )}\int\limits_{0}^{t}\frac{u^{(m)}(\tau )}{%
(t-\tau )^{\alpha +1-m}}d\tau ,
\end{equation}%
where $\;m-1<\alpha <m,m\in \overline{1,2}$.

\section{LINEAR STABILITY ANALYSIS}

Due to the property of Caputo derivative the stability of the
steady-state solutions of the system (\ref{1}) corresponding to
homogeneous equilibrium state
$W(u_{0},\mathcal{A}_{0})=0$\label{eq} can be analyzed by
linearization of the system nearby this constant solution $u_{0}=(\overline{u%
}_{1},\overline{u}_{2})^{T}$. The linearization of FRD system
(\ref{1})\ leads to fractional ODEs with right hand side matrix
$F(k)=\left(
\begin{array}{cc}
(a_{11}-k^{2}l_{1}^{2})/\tau _{1} & a_{12}/\tau _{1} \\
a_{21}/\tau _{2} & (a_{22}-k^{2}l_{2}^{2})/\tau _{2}%
\end{array}%
\right) $, diagonal form of which is given by eigenvalues $\lambda _{1,2}=%
\frac{1}{2}(trF\pm \sqrt{tr^{2}F-4\det F})$ (coefficients $a_{ij}$
represent Jacoby matrix).

For $\alpha :0<\alpha <2$\thinspace\ for every point inside the parabola $%
\det F=tr^{2}F/4$, we can introduce a marginal value $\alpha :\alpha _{0}=$ $%
\frac{2}{\pi }|Arg(\lambda _{i})|$ given by the formula
\cite{gd07,gd08}
\begin{equation}
\alpha _{0}=\left\{
\begin{array}{cc}
\frac{2}{\pi }\arctan \sqrt{4\det F/tr^{2}F-1}, & trF>0, \\
2-\frac{2}{\pi }\arctan \sqrt{4\det F/tr^{2}F-1}, & trF<0.%
\end{array}%
\right.  \label{al}
\end{equation}%
The value of $\alpha $ is a certain additional bifurcation
parameter which
switches the stable and unstable states of the system. At lower $\alpha :$ $%
\alpha <\alpha _{0}=$ $\frac{2}{\pi }|Arg(\lambda _{i})|$, the
system has oscillatory modes, but they are stable. Increasing the
value of $\alpha
>\alpha _{0}=$ $\frac{2}{\pi }|Arg(\lambda _{i})|$ leads to oscillatory
instability.

It is widely known for integer time derivatives \cite{pr,ch,KO}\
that system (\ref{1}) becomes unstable according to either Hopf
($k=0$)
\begin{equation}
trF>0,\quad \det F(0)>0.  \label{ho}
\end{equation}%
or Turing ($k_{0}\neq 0$)\ bifurcations

\begin{equation}
trF<0,\quad \det F(0)>0,\quad \det F(k_{0})<0.  \label{tu}
\end{equation}%
and these both types of instabilities are realized for positive feedback ($%
a_{11}>0)$ \cite{pr,ch,KO}$.$

In the case of fractional derivative index, Hopf bifurcation is
not
connected with the condition $a_{11}>0$ and can hold at a certain value of $%
\alpha $ when the fractional derivative index is sufficiently large \cite%
{cam}.\ Moreover, in fractional RD systems at $\alpha >1$ when it
is easier to satisfy conditions of Hopf bifurcation, we meet a new
type of instability \cite{gd07,gd08}
\begin{equation}
trF<0,\quad 4\det F(0)<tr^{2}F(0),\quad 4\det
F(k_{0})>tr^{2}F(k_{0}). \label{gd}
\end{equation}%
It is worth to analyze inequalities (\ref{gd}) in detail. Taking
into account explicit form of $F(k)$\ the last two conditions can
be rewritten as:
\begin{equation}
(a_{11}\tau _{1}-a_{22}\tau _{2})^{2}>-4a_{12}a_{21}\tau _{1}\tau
_{2}, \label{nho}
\end{equation}%
\begin{equation}
-4a_{12}a_{21}\tau _{1}\tau _{2}>\left[
(a_{11}-k^{2}l_{1}^{2})\tau _{2}-(a_{22}-k^{2}l_{2}^{2})\tau
_{1}\right] ^{2}.  \label{s2}
\end{equation}%
The simplest way to satisfy the last condition is to estimate the
optimal value of $k=k_{0}$
\begin{equation}
k_{0}=2\left( \frac{-a_{12}a_{21}}{l_{1}^{2}/\tau _{2}-l_{2}^{2}/\tau _{1}}%
\right) ^{1/2}.  \label{k0}
\end{equation}%
Having obtained (\ref{k0}), we can estimate the marginal value of
$\alpha _{0}$
\begin{equation}
\alpha _{0}=2-\frac{2}{\pi }\arctan T,  \label{rad}
\end{equation}%
where the expression $T$ has the following view
\begin{equation}
T=\frac{\left( -4a_{12}a_{21}\tau _{1}\tau _{2}\right)
^{1/2}}{\left\vert \left( a_{11}\tau _{2}-a_{22}\tau _{1}\right)
\frac{l_{1}^{2}\tau _{2}+l_{2}^{2}\tau _{1}}{l_{2}^{2}\tau
_{2}-l_{2}^{2}\tau _{1}}\right\vert -a_{11}\tau _{2}-a_{22}\tau
_{1}}.  \label{tt}
\end{equation}

The analysis of expressions (\ref{gd}) shows that at $k=0$ we have
two real eigenvalues that are less than zero, and the system is
certainly stable for the Hopf bifurcation. If the last inequality
takes place for a certain value of $k_{0}\neq 0,$ we can get two
complex eigenvalues, and a new type of instability, connected with
the interplay between the determinant and the trace of $F(k)$ of
the linearized system, emerges. With such type of eigenvalues, it
is possible to determine the value of fractional derivative index
when the system becomes unstable for Hopf bifurcation with this
wave number \cite{gd07}.

\section{ FRACTIONAL REACTION DIFFUSION SYSTEM WITH CUBIC NONLINEARITY}

To demonstrate the properties of FRD system, let us consider the
model with cubic dependance for activator variable
$W_{1}=u_{1}-u_{1}^{3}-u_{2}$ and
the linear for the inhibitor variable $W_{2}=-u_{2}+\beta u_{1}+\mathcal{A}$%
. This model was proposed firstly by R. FitzHugh \cite{fh} for
description of the propagation of voltage impulse through a nerve
axon and is known as Bonhoeffer-van der Pol model. In RD systems
this model was considered in many books and articles (see for
example  \cite{pr,ch,KO}). The homogeneous solution of variables
$\overline{u}_{1}$ and $\overline{u}_{2}$ can be obtained from the
system of equations $W_{1}(\overline{u}_{1},\overline{u}_{2})=0,W_{2}(%
\overline{u}_{1},\overline{u}_{2})=0$, which in their turn
determine two nullclines $\overline{u}_{2}(\overline{u}_{1})$. The
intersection of these nullclines in the point
$P=(\overline{u}_{1},\overline{u}_{2})$ is determined by equation:
$\overline{u}_{1}-\overline{u}_{1}^{3}-\beta
\overline{u}_{1}-\mathcal{A=}0.$ In this case the values of
external parameters $A$,$\beta $ determine the value of
$\overline{u}_{1}$\ and this makes it possible to investigate the
conditions of different types of instability explicitly
considering parameter $\overline{u}_{1}$ as the main parameter for
system analysis.

For investigation of the Hopf bifurcation let us consider
homogeneous perturbation with $k=0$. The linear analysis of the
system with $\alpha =1$ shows that, if $\tau _{1}/\tau _{2}>1$,
the solution corresponding to any intersections of two isoclines
is stable. The smaller is the ratio of $\tau _{1}/\tau _{2}$, the
wider is the instability region (solid lines with points on Fig.
1). Formally, at $\tau _{1}/\tau _{2}\rightarrow 0$, the
instability region for $\overline{u}_{1}$\ coincides with the
interval ($-1,1 $) where the null cline $W(u_{1},u_{2})=0$ has its
increasing part (Fig. 2b). These results are very widely known in
the theory of nonlinear dynamical systems \cite{pr,ch,KO}.

In the fractional differential equations the conditions of the
instability depend on the value of $\alpha $ and we have to
analyze the real and the imaginary parts of the existing complex
eigenvalues, especially the equation:
\begin{equation}
4\det F-tr^{2}F=\frac{4((\beta -1)+\overline{u}_{1}^{2})}{\tau _{1}\tau _{2}}%
-\left( \frac{(1-\overline{u}_{1}^{2})}{\tau _{1}}-\frac{1}{\tau _{2}}%
\right) ^{2}>0.  \label{ri}
\end{equation}%
In fact, with complex eigenvalues, it is possible to find the
corresponding value of $\alpha $ where the condition $\alpha
>\alpha _{0}$ is true. Omitting simple calculation, we can write
an equation for marginal values of $\overline{u}_{1}$
\begin{equation}
\overline{u}_{1}^{4}-2(1+\frac{\tau _{1}}{\tau _{2}})\overline{u}_{1}^{2}+%
\frac{\tau _{1}^{2}}{\tau _{2}^{2}}-2\frac{\tau _{1}}{\tau
_{2}}(2\beta -1)+1=0,  \label{ri1}
\end{equation}%
and solution of this biquadratic equation gives us the domain
where the oscillatory instability can arise
\begin{equation}
\overline{u}_{1}^{2}=1+\frac{\tau _{1}}{\tau _{2}}\pm 2\sqrt{\beta \frac{%
\tau _{1}}{\tau _{2}}}.  \label{t}
\end{equation}%
This expression estimates the maximum and minimum values of $\overline{u}%
_{1} $ where the system can be unstable at marginal value of
$\alpha =\alpha _{0}=2$ as a function of $\tau _{1}/\tau _{2}$ and
$\beta $.

The typical stability domains for considered FRD system in the coordinates ($%
\overline{u}_{1},\tau _{1}/\tau _{2}$) for different values of
fractional derivative index $\alpha $\ are presented in Fig.1a-d,
where curves corresponding to $\alpha =1$\ are denoted by more
solid lines. This makes it possible to see how other curves
$\alpha \neq 1$ are located with respect to standard system
$\alpha =1$. The solid points on both sides denote the interval of
maximum instability. For each particular value $\alpha $ in the
region between the corresponding curve and horizontal axis, the
system is unstable with wave numbers $k=0$, and outside it is
stable. Figures (a) and (b) present the plots for the value of
$\beta =1.01$. The left-hand side
plot corresponds to $\alpha <1$ and the right-hand side plot corresponds to $%
\alpha >1$. It is easy to see from the plot (a) that if the value
of $\tau _{1}/\tau _{2}$ increases (we move along vertical axis)
the instability
domain decreases ($\alpha <1$). At $\tau _{1}/\tau _{2}=1$ \ and $\alpha =1$%
\ it vanishes completely. The solid point in the middle of each
plots
corresponds to the case when the system becomes stable for all values of $%
\overline{u}_{1}:$ point $(0,1)$ in coordinates
($\overline{u}_{1},\tau _{1}/\tau _{2}$).\ The situation changes
for $\alpha >1.$ The system is unstable not only for $\tau
_{1}/\tau _{2}<1$\ but also for $\tau _{1}/\tau _{2}>1.$ An
increase in $\alpha $ makes the instability domain much wider with
respect to two coordinates ($\overline{u}_{1},\tau _{1}/\tau
_{2}$) and we obtain butterfly like domains for $\alpha >1.$ This
means that with increasing $\tau _{1}/\tau _{2}$\ (moving along
vertical axis) the system becomes stable in the center and
unstable at greater values $\overline{u}_{1} $ In such case the
instability domain becomes symmetric along vertical axis with the
minimum point at the $\overline{u}_{1}=0$. We observe similar
behavior for large values of $\beta .$ Instability domains in Fig.
1c,d
are presented for $\beta =10$ and show the same trend with respect to $%
\alpha ,$ but the region for $\alpha >1$ is much greater than for
$\beta =1.01$. At the same time, for $\alpha <1$ and $\beta =10$
instability domain
shrinks very sharply in comparison to the same domain plot when $\beta =1.01$%
.

\begin{figure}[tbp]
\begin{center}
\begin{tabular}{cc}
\includegraphics[width=0.24\textwidth]{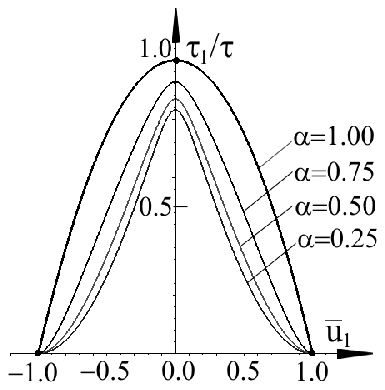} & %
\includegraphics[width=0.24\textwidth]{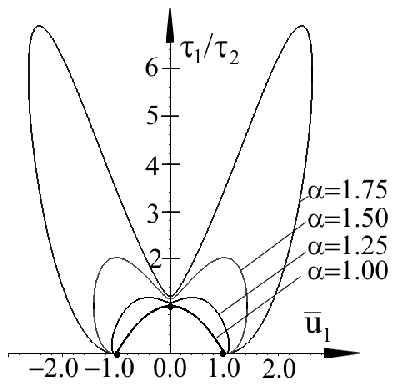} \\
(a) & (b) \\
\includegraphics[width=0.24\textwidth]{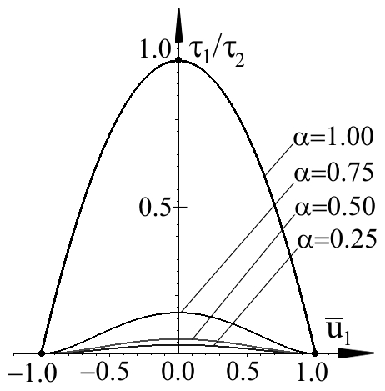} & %
\includegraphics[width=0.24\textwidth]{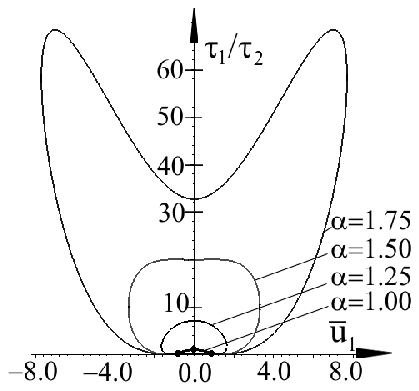} \\
(c) & (d) \\
&
\end{tabular}%
\end{center}
\caption{Instability domains in coordinates $(\overline{u}_{1},\protect\tau %
_{1}/\protect\tau _{2})$ for a fractional order reaction-diffusion
system with sources $W_{1}=u_{1}-u_{1}^{3}-u_{2}$,
$W_{2}=-u_{2}+\beta u_{1}+\mathcal{A}$
for different values of $\protect\alpha _{0}=0.25,0.5,0.75,1.0,1.25,1.5,1.75$%
. The results of computer simulation obtained at $l_{1}=l_{2}=0$ for: $%
\protect\beta =1.01$ - (a), (b) and $\protect\beta =10.0$ - (c),
(d) . } \label{rys1}
\end{figure}

It is possible to obtain solid understanding of the mechanism of
the instability from the plot of eigenvalues. Typical instability
domains for the same parameters as on Fig. 1a,b for $k=0$ are
presented on the Fig. 2a. Horizontal lines (i, ii, iii) on the
instability domain plot correspond to eigenvalues plots below. Let
us analyze each of the possible situations in more detail.

For the case (i) we have sub-domains with real positive, real
negative and complex eigenvalues. The easiest way of obtaining
instability is realized at
$|\overline{u}_{1}|<\overline{u}_{1}^{E}$ when all the roots are
real and positive (Fig. 2a(i)). This region is presented by dark
grey color and positive eigenvalues mean that the system is
unstable practically for any
value of $\alpha >0.$ Inside the domain $|\overline{u}_{1}^{E}|<|\overline{u}%
_{1}|<|\overline{u}_{1}^{C}|$ there is a certain domain of $\alpha
:(0<\alpha <2)$ where the Hopf bifurcation takes place. Point $D$
divides the region into two domains where Re$\lambda <0$\ and
Re$\lambda >0.$ In the domain Re$\lambda <0$ the system could be
unstable according to greater values of $\alpha >1.$ In turn, for
Re$\lambda >0,$ the system could be stable at $\alpha <1.$ In
other words, between points $C$ and $E$ we have eigenvalues with
imaginary part, and the value of $\alpha $ can change the
stability of the FRD system. In the \ domain $|\overline{u}_{1}|>|\overline{u%
}_{1}^{C}|$ we have two real and negative roots and, as a result,
the system is stable.
\begin{figure}[tbp]
\begin{center}
\begin{tabular}{c}
\includegraphics[width=0.29\textwidth]{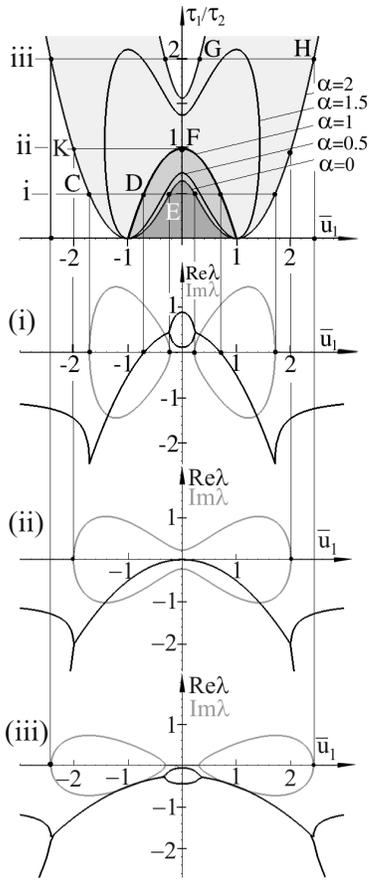} \\

\end{tabular}%
\end{center}
\caption{ Instability domains ($Re\protect\lambda $ - black lines,
$Im\protect\lambda $ - grey lines) for $k=0,\protect\beta =1.05$
and different proportions of $\protect\tau _{1}/\protect\tau %
_{2}=0.5-(i),1.0-(ii),2.0-(iii)$.} \label{rys}
\end{figure}

\begin{figure}[tbp]
\begin{center}
\begin{tabular}{c}

\includegraphics[width=0.29\textwidth]{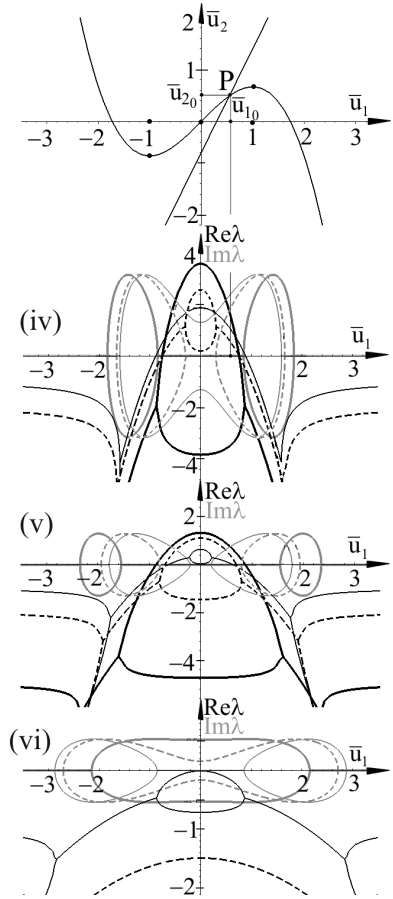} \\

\end{tabular}%
\end{center}
\caption{  The null-clines for $%
\protect\beta =2.1,A=0.5$ and eigenvalues for different values of
$k$ ($k=0$ - hair-lines, $k=1$ - dash lines, $k=2$ - thick lines)
-\textbf{(b)}. The
eigenvalues are presented for the following parameters: $l_{1}=0.025,\protect%
\beta =2.1,\protect\tau _{1}/\protect\tau _{2}=0.21$ - (iv), $l_{1}=0.1,%
\protect\beta =1.01,\protect\tau _{1}/\protect\tau _{2}=0.6$ - (v), $%
l_{1}=2.1,\protect\beta =1.01,\protect\tau _{1}/\protect\tau
_{2}=3.5$ - (vi). } \label{rys}
\end{figure}

For system parameters corresponding to the case (ii) the real part
of eigenvalues becomes less than zero for all $\overline{u}_{1}$.
At the same time, for $|\overline{u}_{1}|<|\overline{u}_{1}^{K}|$
the roots are complex
and and according to condition (\ref{al}) instability takes place for $%
\alpha >\alpha _{0}>1.$ For
$|\overline{u}_{1}|>|\overline{u}_{1}^{K}|$ the roots become real
and negative, and the system is stable.

In the case (iii) at the center of $\overline{u}_{1}$ ( $|\overline{u}_{1}|<|%
\overline{u}_{1}^{G}|)$ we have two real negative eigenvalues, and
the
system is stable. For $|\overline{u}_{1}^{G}|<|\overline{u}_{1}|<|\overline{u%
}_{1}^{H}|$ we have complex roots and certainly according to condition (\ref%
{al}) instability takes place for $\alpha >\alpha _{0}>1.$\ In
this case, the instability domain consists of two symmetrical
regions separated by a stable region at the center where the
system is stable for any $\alpha $. For
$|\overline{u}_{1}|>\overline{u}_{1}^{H}$ the system is stable
again.

Let us analyze the Turing Bifurcation ($k\neq 0$). Eigenvalues for
different values of $k$ are presented in Fig. 2b. The top plot
corresponds to nullclines of the systems just to show that
nullcline intersection determines eigenvalues in Fig. 2b(iv). We
present eigenvalues for $k=1$ and $k=2$ and for comparison $k=0$.
It can be seen from the picture (iv)
that at intersection of nullclines in the vicinity of zero value of $%
\overline{u}_{1}$\ nonhomogeneous modes have much greater values
and we can
expect a formation of stationary dissipative structures. If the ratio $%
l_{1}/l_{2}$ is sufficiently small, Turing bifurcation is dominant
for all region $\overline{u}_{1}<1.$ Analyzing (\ref{tu}) we can
conclude that these conditions are practically the same for
fractional and standard RD systems. However, what is very
important is that the transient processes and the dynamics of
these systems are different, and for this reason final attractors
can often be different even though the linear conditions of
instability look the same.

Now let us consider that the system parameters are close to the
ones
represented by point $P$ in Fig. 2b. As a result, for certain ratio of $%
Im\lambda $ and $Re\lambda $ we expect the formation of
oscillatory inhomogeneous structures. However, if the solution
$\overline{u}_{1}$ is close to zero, the decrease of $\alpha $
leads to steady state dissipative structures. Such trend is quite
general and if in standard system we have steady state solutions,
the increase of $\alpha $ in FRD\ system leads to non stationary
structures. In this case, by changing intersection point of
nullclines or value of $\alpha $ we can stimulate stationary or
temporary pattern formation. If the absolute value of eigenvalues
for $k=0$ and $k\neq 0 $ are comparable we can expect more complex
spatio-temporal dynamics.

Above we have considered that the linearized system is unstable
for either Hopf or Turing bifurcation. Below, we consider the case
when we don't have
Turing or Hopf bifurcation. For realization of instability conditions (\ref%
{gd}) for\textbf{\ }$k\neq 0,Im\lambda \neq 0$\textbf{\ }the
fractional derivative index must be greater than some critical
value $\alpha _{0}$. Eigenvalues for such instability are
presented in Fig. 2b(v,vi). We can see that outside a small domain
in the center the system is stable for  $k=0$. At the same time on
this interval we have complex eigenvalues for $k\neq 0$\ (Fig. 2b
(v)). In the plot (v) for $l_{1}/l_{2}<1$ we have a separate
domain for $k=2$ where we can expect inhomogeneous oscillations
with this wave number.

In Figure (vi) we present the situation where all roots have
$Re\lambda <0$
and in standard system we do not have instability at any values of $%
\overline{u}_{1}.$\ In FRD system the roots are complex for select
values of $k$ ($k\neq 0$). In other words, for such system at
$\alpha >1$ we can obtain conditions of Hopf bifurcation
(\ref{gd}) which lead to inhomogeneous oscillatory structures
\cite{gd07,gd08} even for $l_{1}/l_{2}>2$. This situation can be
predicted from symmetrical view of expression (\ref{tt}) for the
system under consideration

\begin{equation}
T=2\sqrt{\beta }/\left[ \left\vert \left(
1-\overline{u}_{1}^{2}\right) \frac{l_{1}^{2}\tau
_{2}+l_{2}^{2}\tau _{1}}{l_{1}^{2}\tau _{2}-l_{2}^{2}\tau
_{1}}\right\vert -\left( 1-\overline{u}_{1}^{2}\right) \tau
_{2}/\tau _{1}+1\right] .  \label{tt1}
\end{equation}%
The plot of these surfaces, as a function of $l_{1}/l_{2}$ and
$\tau
_{2}/\tau _{1,}$ are presented in plots 3 (a-d) for different values of $%
\overline{u}_{1}.$\ We can see that in Fig. (a) and (b) the
maximum value of $T$ is reached at the boundary and
$l_{1}/l_{2}<<1$ (Fig. 3a) and $\tau _{2}/\tau _{1}<<1$ (Fig. 3b).
For figure. 3 c,d the optimal instability conditions are reached
at certain combination of the parameters $l_{1}/l_{2}$ and $\tau
_{1}/\tau _{2}$ and we can expect inhomogeneous oscillations at
different relationships of $l_{1}/l_{2}$ and $\tau _{1}/\tau
_{2}.$

\begin{figure}[tbp]
\begin{center}
\begin{tabular}{cc}
\includegraphics[width=0.24\textwidth]{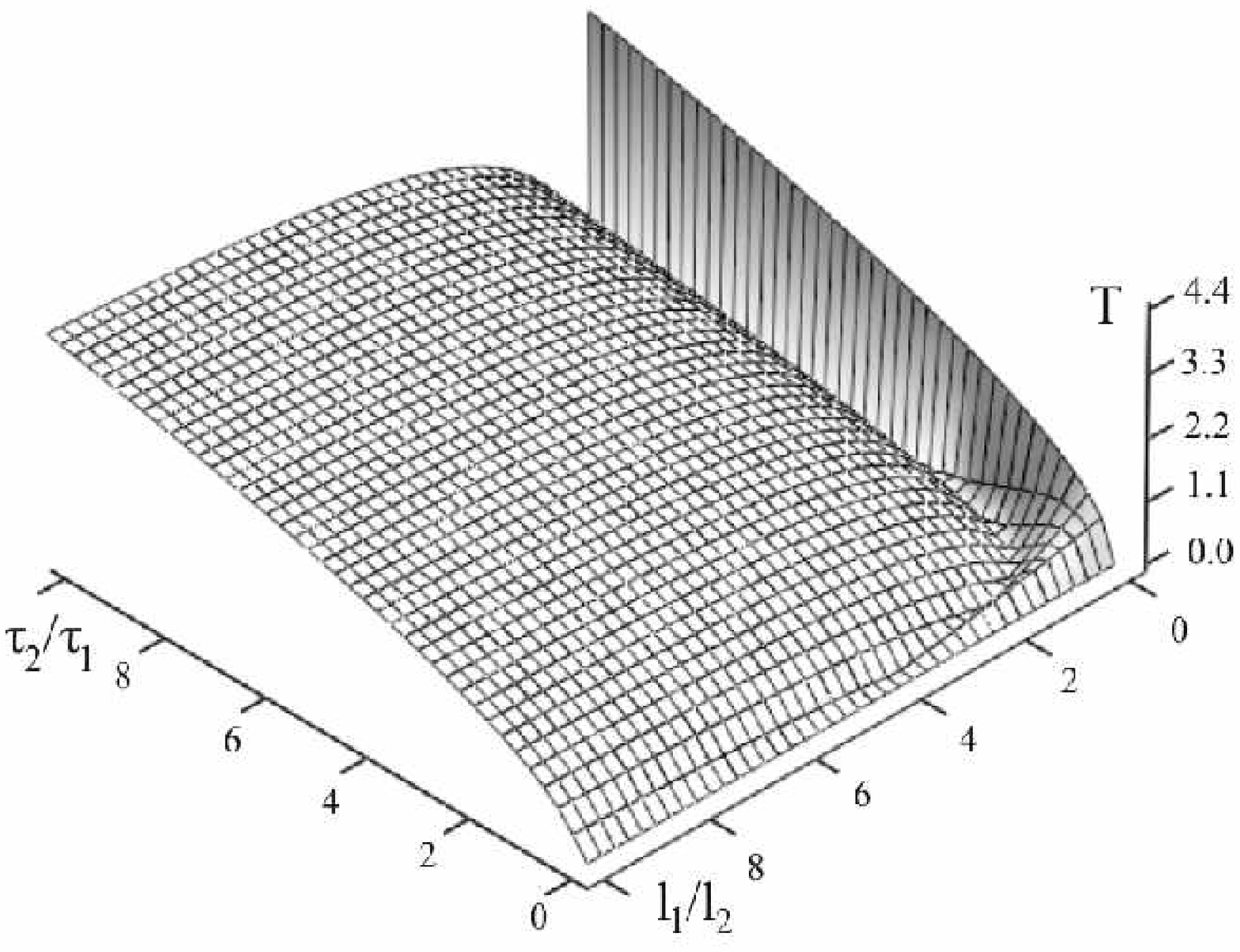} & %
\includegraphics[width=0.24\textwidth]{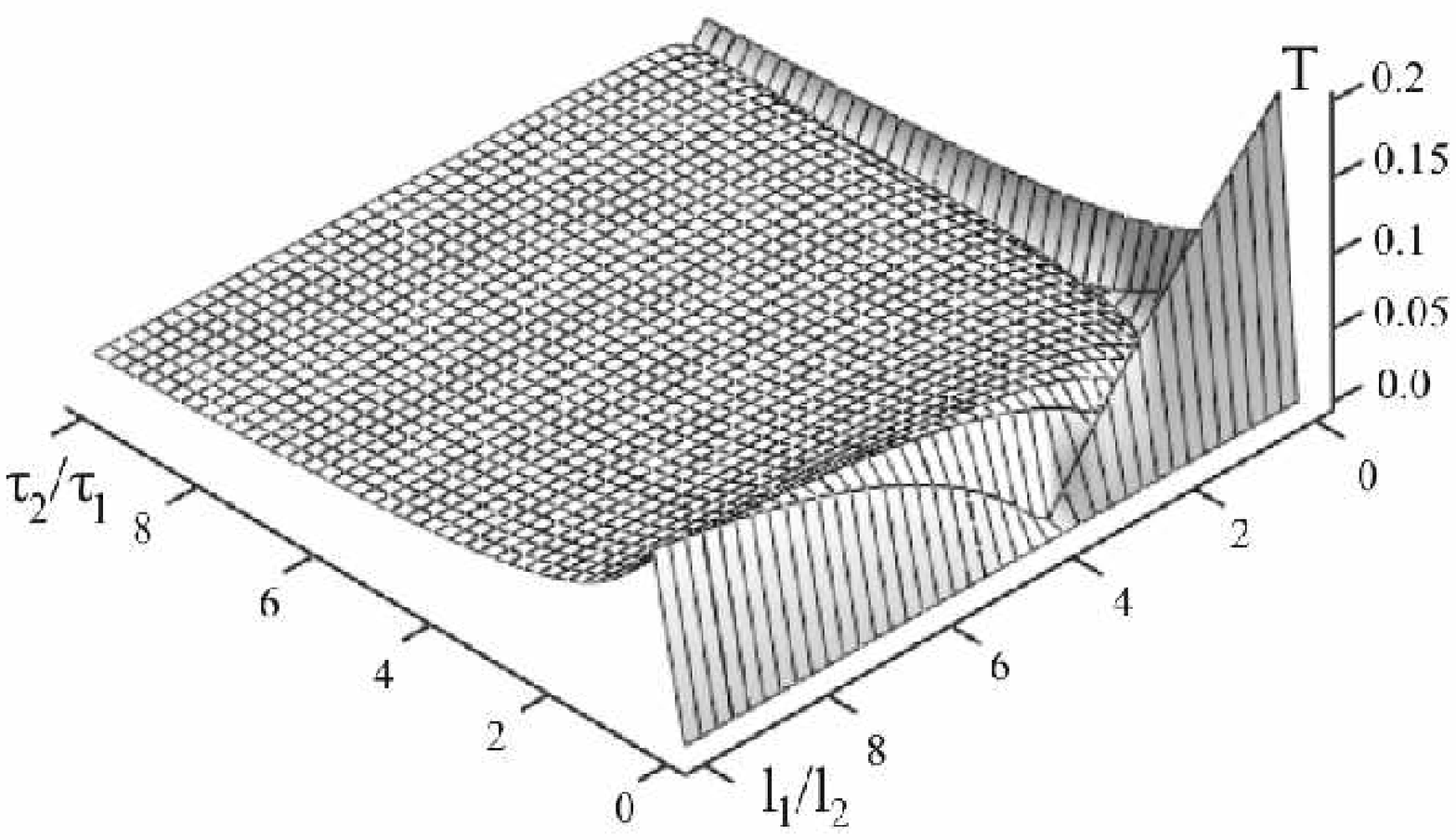} \\
(a) & (b) \\
\includegraphics[width=0.24\textwidth]{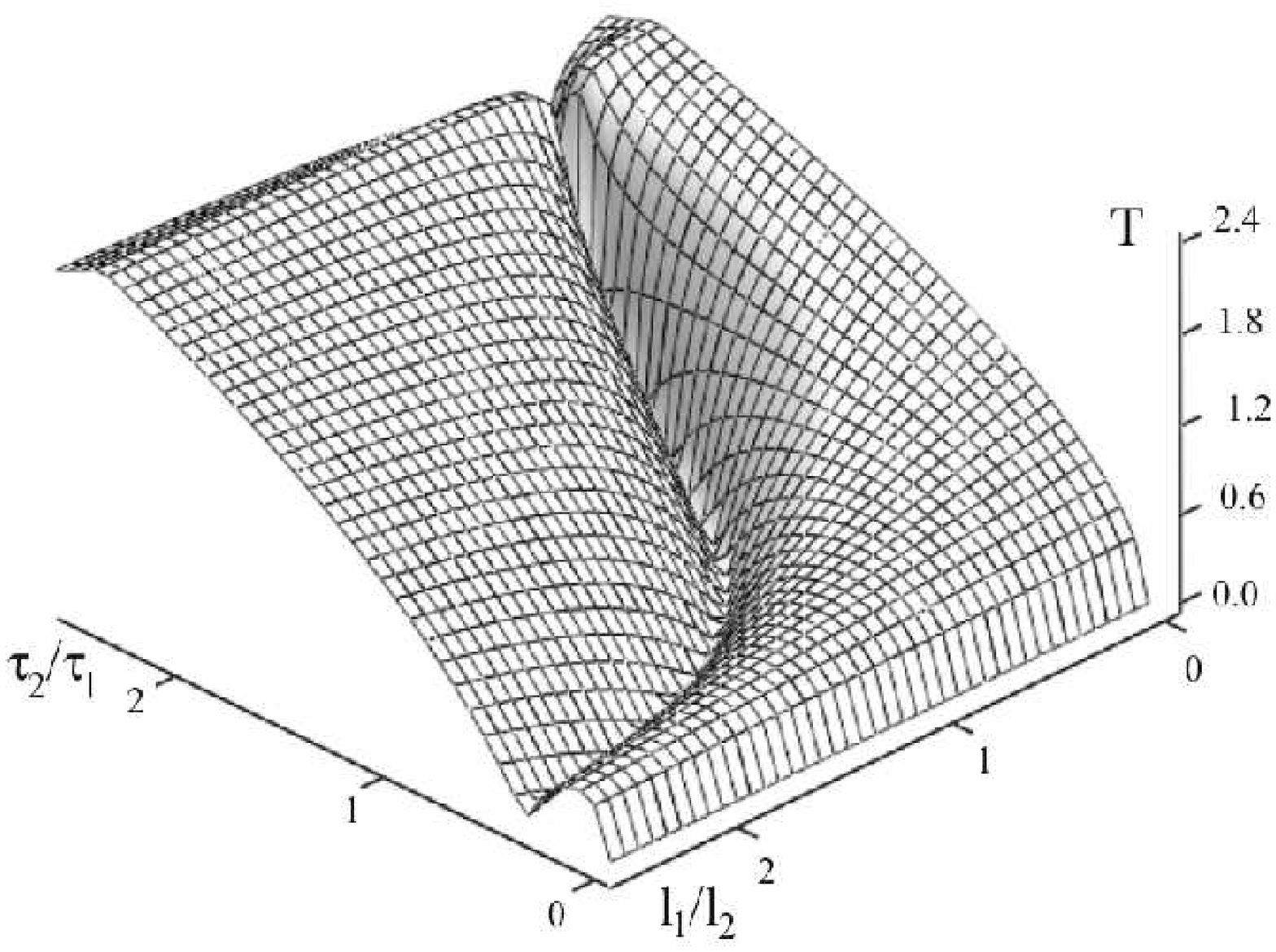} & %
\includegraphics[width=0.24\textwidth]{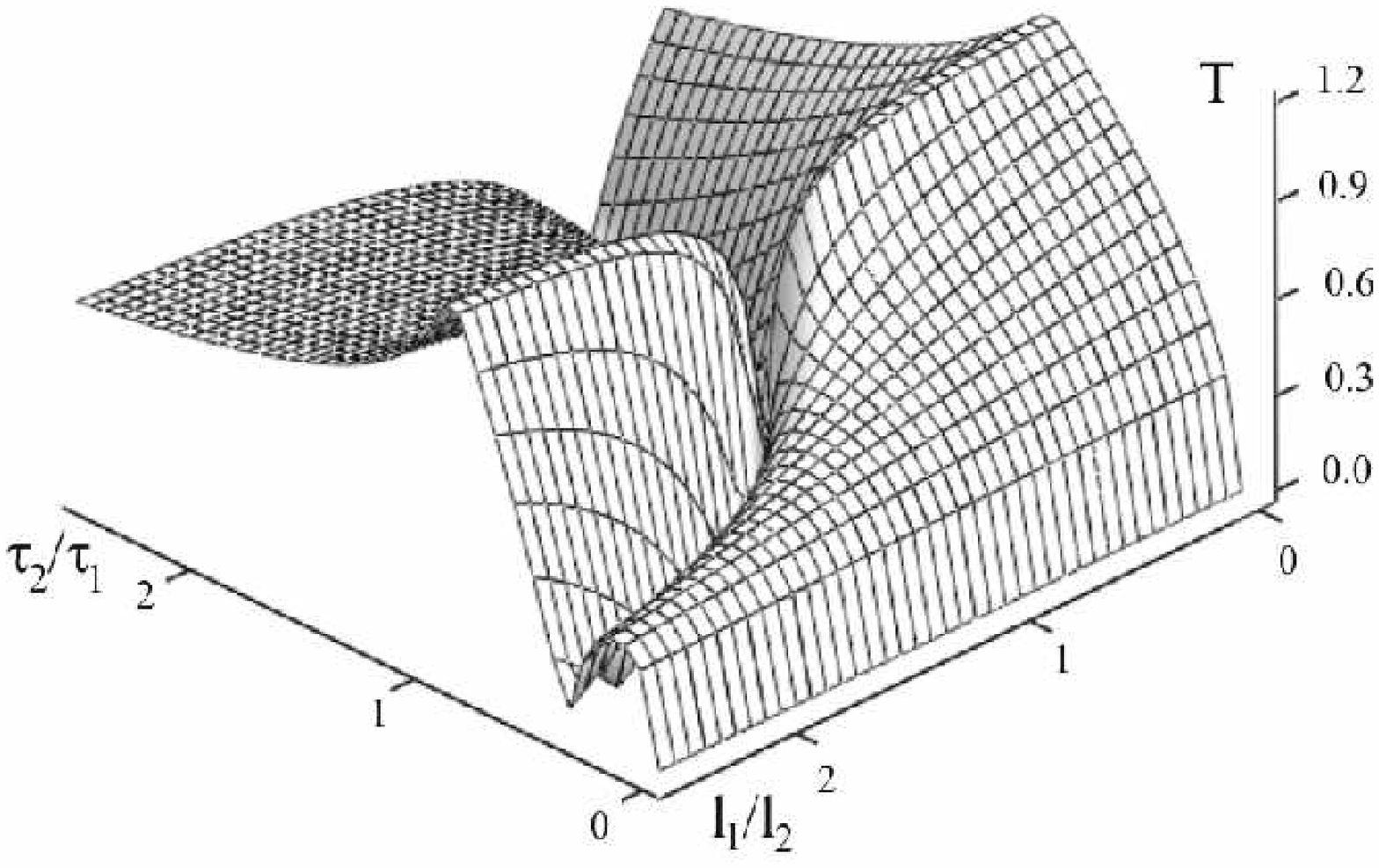} \\
(c) & (d) \\
&
\end{tabular}%
\end{center}
\caption{The view of the surface $T$ in coordinates $\quad (\l _{1}/l_{2},%
\protect\tau _{2}/\protect\tau _{1})\quad $ for $\protect\beta =2$
and
different values of $u_{1}$ ($u_{1}=0.1$ - (a), $u_{1}=5.0$ - (b), $%
u_{1}=1.25$ - (c), $u_{1}=1.5$ - (d)) } \label{rys10}
\end{figure}

We expect the oscillatory structures to emerge if at given values of $%
\overline{u}_{1}$, $l_{1}/l_{2}$ and $\tau _{1}/\tau _{2}$, the
fractional derivative index $\alpha $ is greater than the one
represented on the surface and less than the one needed for Hopf
bifurcations for $k=0$. This means that only the perturbations
with these wave numbers are unstable, and they are unstable for
oscillatory fluctuations. This situation is
qualitatively different from the integer RD system whether either Turing ($%
k\neq 0$) or Hopf bifurcation ($k=0$) takes place, and this
depends on which condition is easier to realize. Thus, in the
system under consideration, we can choose the parameter when we
don't have Turing and Hopf bifurcations (for $k=0$)\ at all.
Nevertheless, we obtain that conditions for Hopf bifurcation can
be realized for a nonhomogeneous wave number. As it is seen from
the figure, there are conditions where only instability according
to non-homogeneous wave numbers holds. As a result, perturbations
with $k=0$ relax to the homogenous state, and only the
perturbations with a certain value of $k$ become unstable and the
system exhibits inhomogeneous oscillations.

\section{Pattern Formation}

The results of the numerical simulation of the fractional RDS
(\ref{1}) are presented on Fig. 4,5. From the pictures we can see
that in such system we obtain a rich scenario of pattern
formation: standard homogeneous oscillations, Turing stable
structures, interacting inhomogeneous structures and inhomogeneous
oscillatory structures. We have obtained that the ratio of
characteristic times and the order of fractional derivative
qualitative transform pattern formation dynamics: homogeneous
oscillations in the first limiting case and stationary dissipative
structures in the second one \cite{cam}. Here we show that the
change in any parameter which qualitatively changes the
eigenvalues of the linearized system can change the system
dynamics. Spatiotemporal dynamics of the FRD system can mainly be
determined by the maximum eigenvalues for the corresponding modes.
In Fig. 4a,b we can see stationary dissipative structures as a
result of formation of the unstable mode presented in Fig. 2b(iv)
for $\alpha <1.$ External parameter $A$ determines the
intersection point and the slope of the isoclines in this point
and the power for each particular mode at this parameter. In
particular, for $A=0.25$ nullclines intersect at the point where
maximum value has eigenvalue with $k=2,$ which is responsible for
Turing bifurcation with stationary structures. For A=0.5 (Fig.
2b(iv)) nullclines intersect at the point where Hopf bifurcation
takes place. The characteristic feature for these two limit cases
is the instantaneous formation of either dissipative structures or
homogeneous oscillations. Increasing influence of Hopf bifurcation
when the Turing one is dominant, or increasing Turing bifurcation
when Hopf is dominant, leads to more complicated transient
dynamics (Fig. 4c,d). When conditions of these two instabilities
practically coincide, we can obtain either oscillatory
inhomogeneous structures or modulated homogeneous oscillations
(Fig. 4e,f). Moreover, at parameters, when the real part of
eigenvalues is close to zero, small variation of $\alpha $ changes
the type of bifurcation. This trend is typical for any $\alpha
\leq 1.$ For $\alpha >1$ the structure formation can be much more
complicated.

\begin{figure}[tbp]
\begin{center}
\begin{tabular}{cc}
\includegraphics[width=0.24\textwidth]{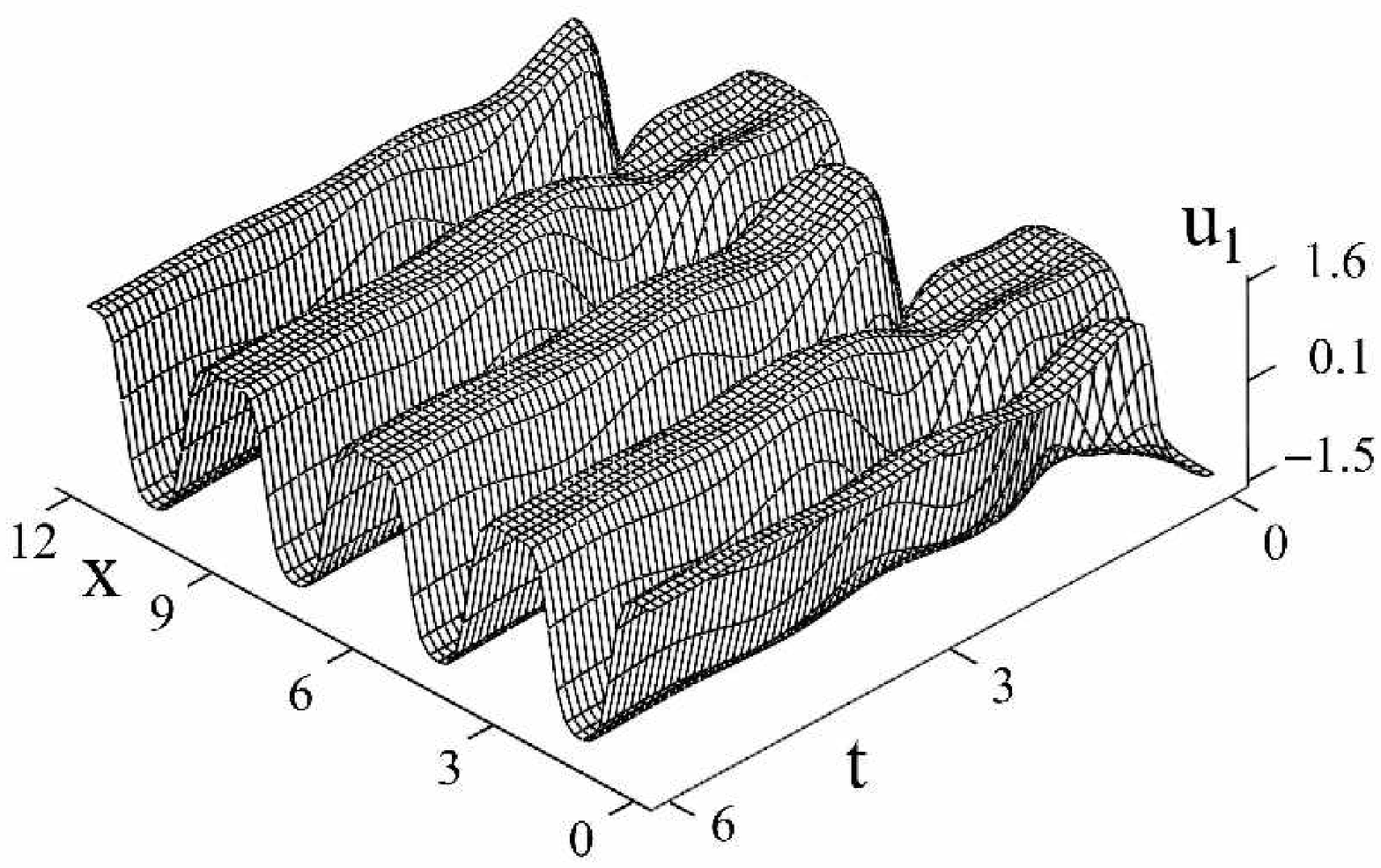}
&
\includegraphics[width=0.24\textwidth]{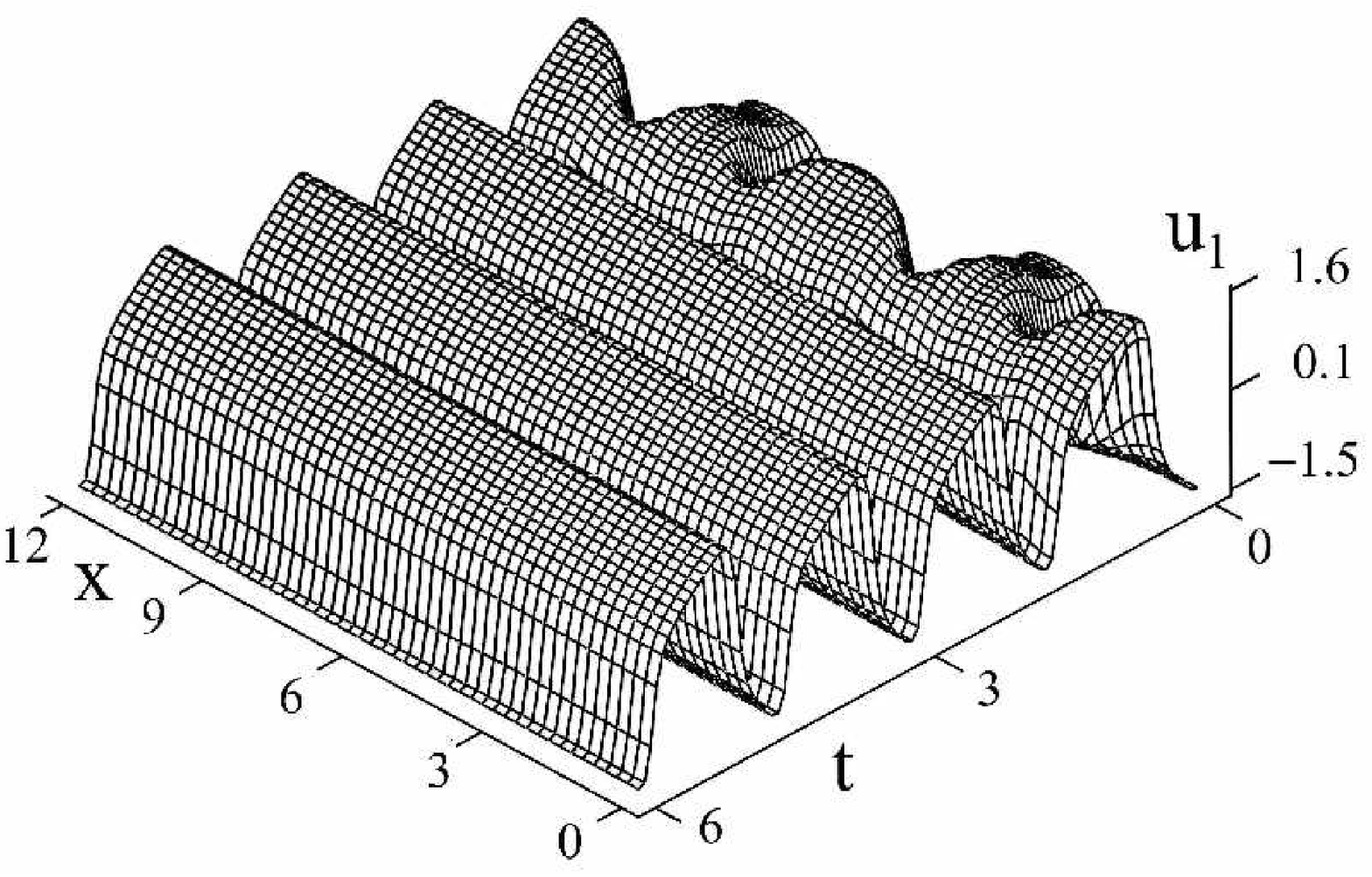}
\\
(a) & (b) \\
    &     \\
\includegraphics[width=0.24\textwidth]{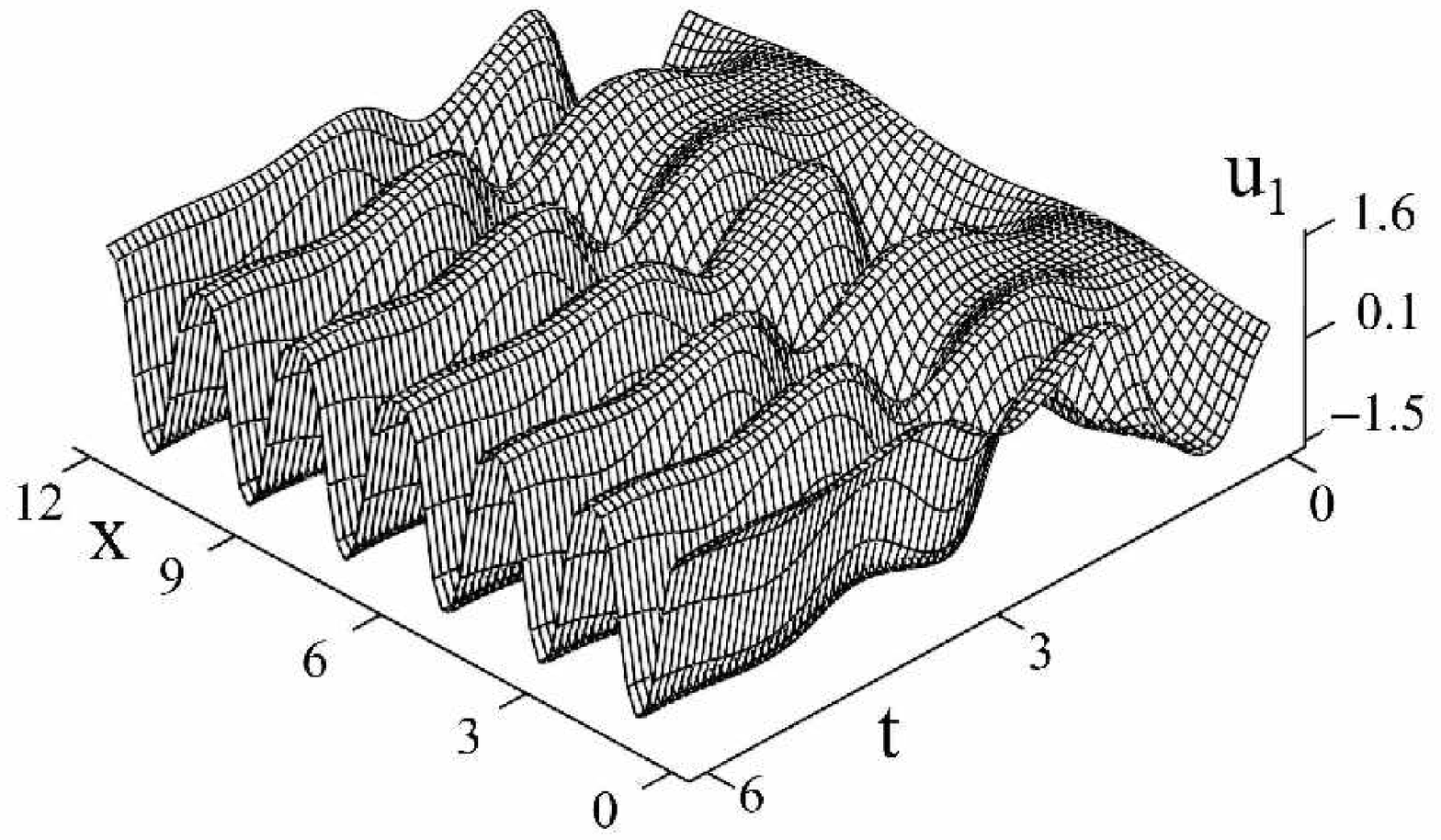}
&
\includegraphics[width=0.24\textwidth]{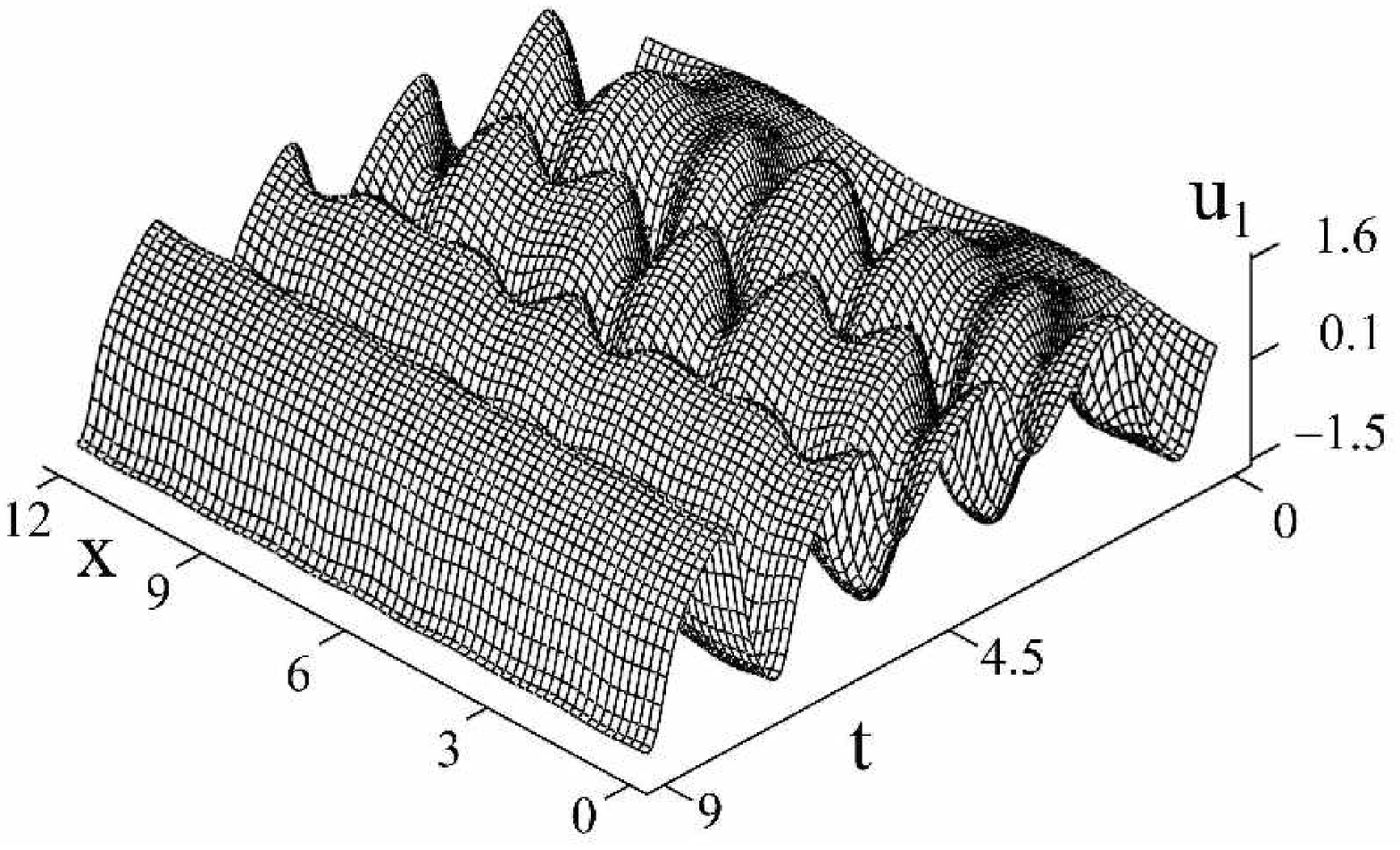}
\\
(c) & (d) \\
    &    \\
\includegraphics[width=0.25\textwidth]{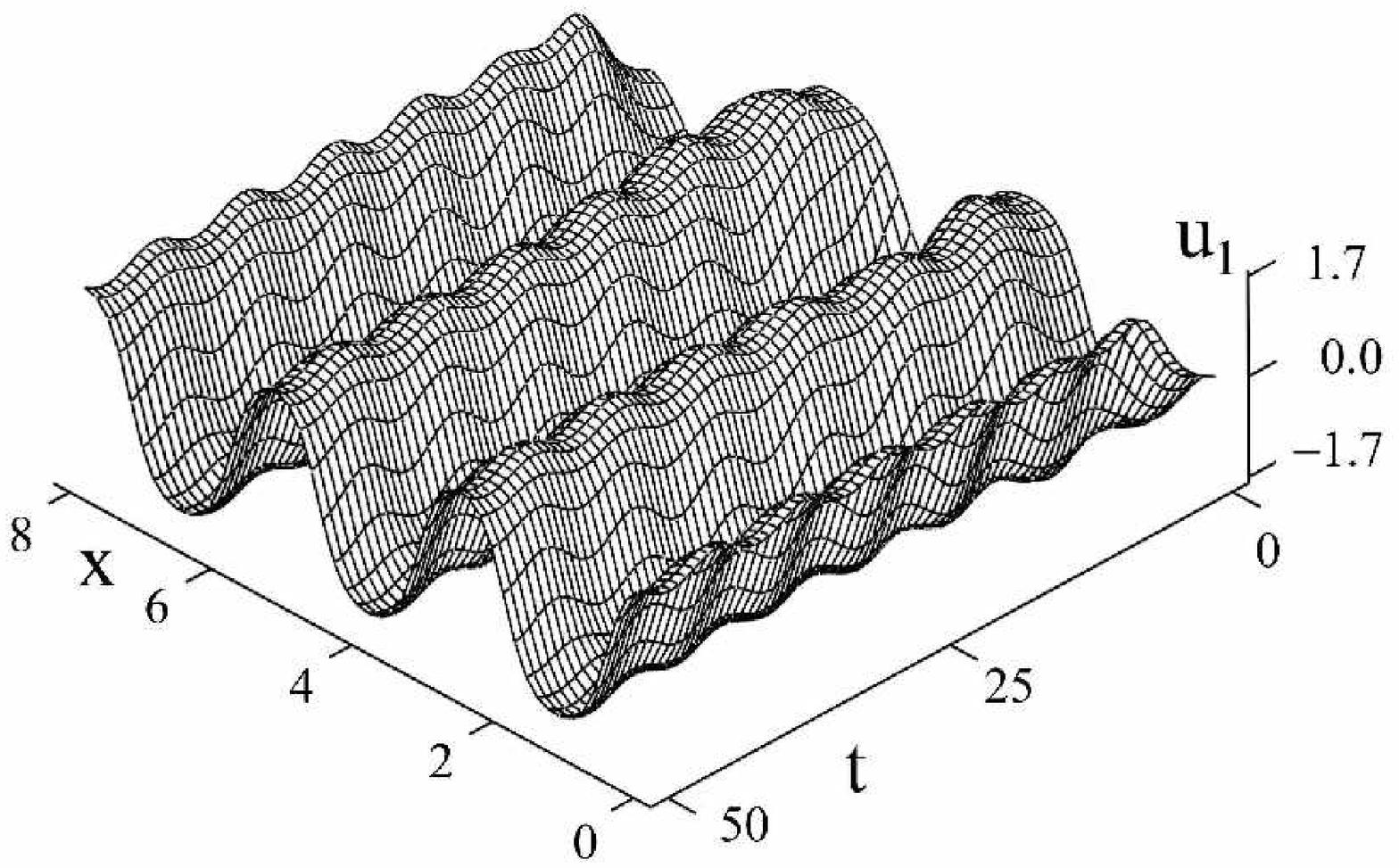}
& \includegraphics[width=0.25%
\textwidth]{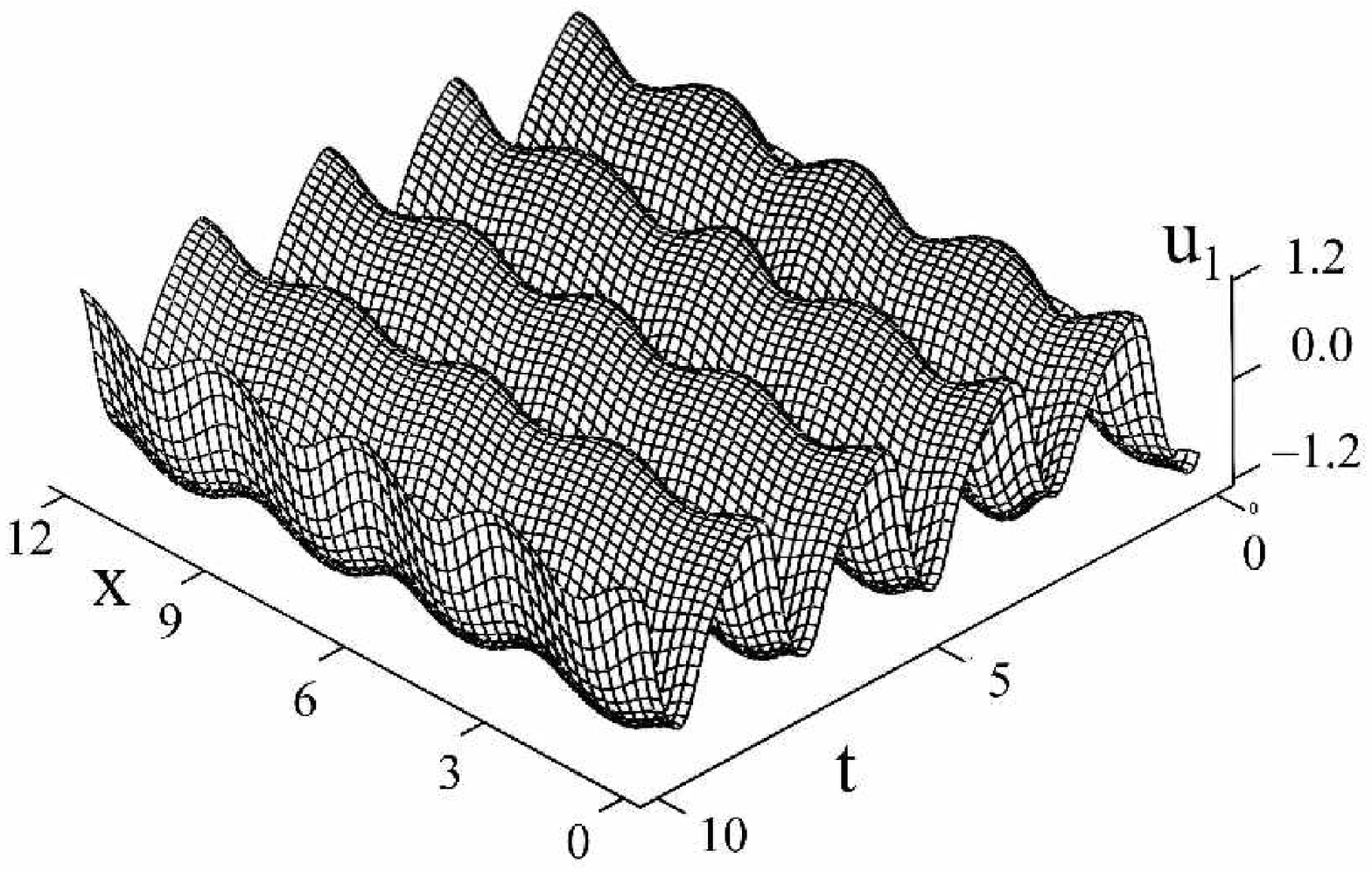} \\
(e) & (f) \\
    &     \\
\end{tabular}%
\end{center}

\caption{Dynamics of pattern formation  for $u_{1}$ variable. The
results of computer
simulations of the system  at parameters: $\protect\alpha =0.8$, $%
A=-0.25 $, $\protect\beta =2.1$, $l_{1}=0.025$, $l_{2}=1$, $\protect\tau %
_{1}/\protect\tau _{2}=0.1$ -- (a); $\protect\alpha =0.8$, $A=-0.55$, $%
\protect\beta =2.1$, $l_{1}=0.025$, $l_{2}=1$, $\protect\tau _{1}/\protect%
\tau _{2}=0.1$ -- (b); $\protect\alpha =0.8$, $A=-0.4$, $\protect\beta =2.1$%
, $l_{1}=0.025$, $l_{2}=1$, $\protect\tau _{1}/\protect\tau
_{2}=0.1$ --
(c); $\protect\alpha =0.8$, $A=-0.45$, $\protect\beta =2.1$, $l_{1}=0.025$, $%
l_{2}=1$, $\protect\tau _{1}/\protect\tau _{2}=0.1$ -- (d); $\protect\alpha %
=1.6$, $A=-0.01$, $\protect\beta =1.05$, $l_{1}=0.05$, $l_{2}=1$, $\protect%
\tau _{1}/\protect\tau _{2}=1.45$ -- (e); $\protect\alpha =0.7$, $A=-0.3$, $%
\protect\beta =2.1$, $l_{1}=0.05$, $l_{2}=1$, $\protect\tau _{1}/\protect%
\tau _{2}=0.2$ -- (f); } \label{rys11}
\end{figure}

\begin{figure}[tbp]
\begin{center}
\begin{tabular}{cc}
\includegraphics[width=0.25\textwidth]{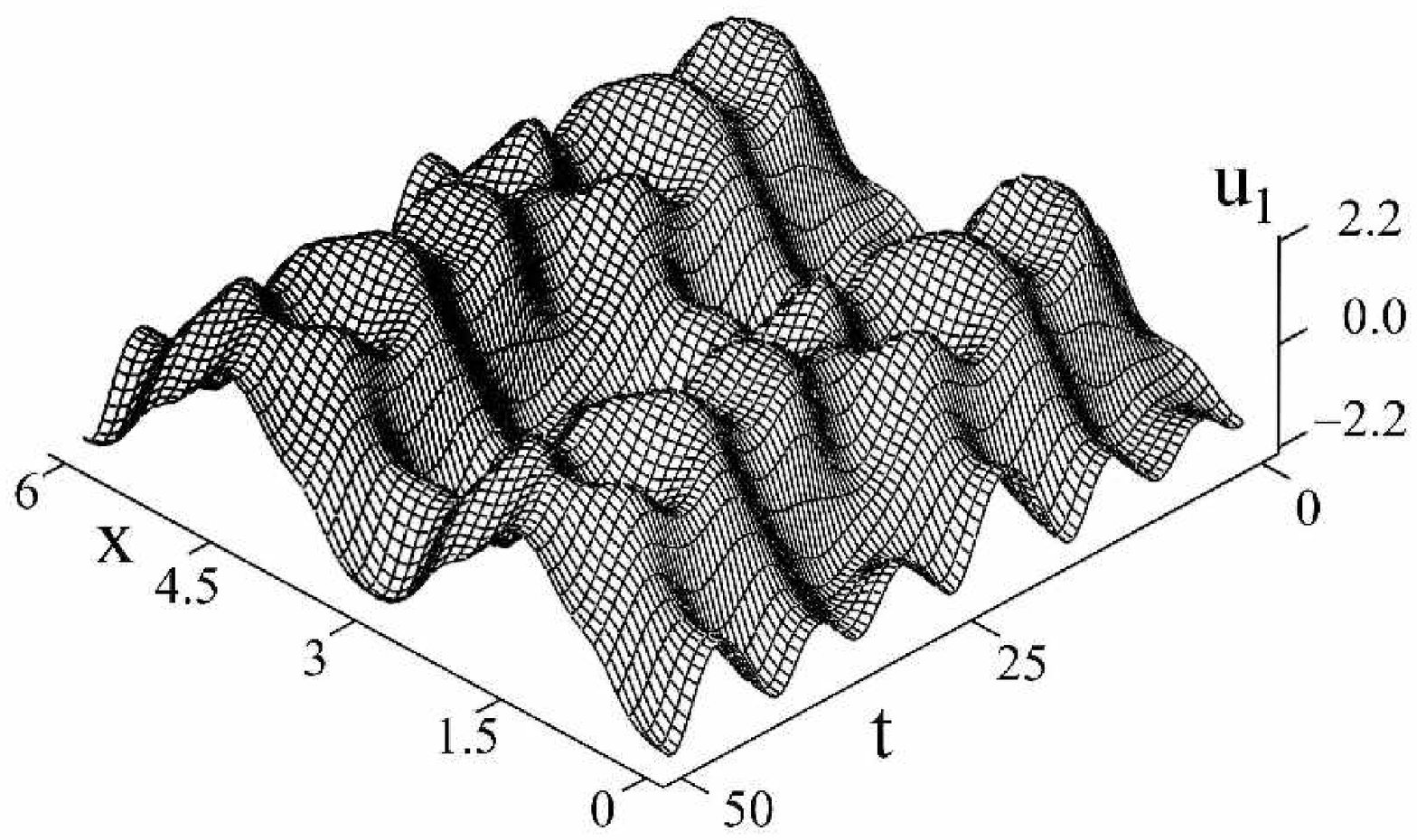}&
\includegraphics[width=0.25\textwidth]{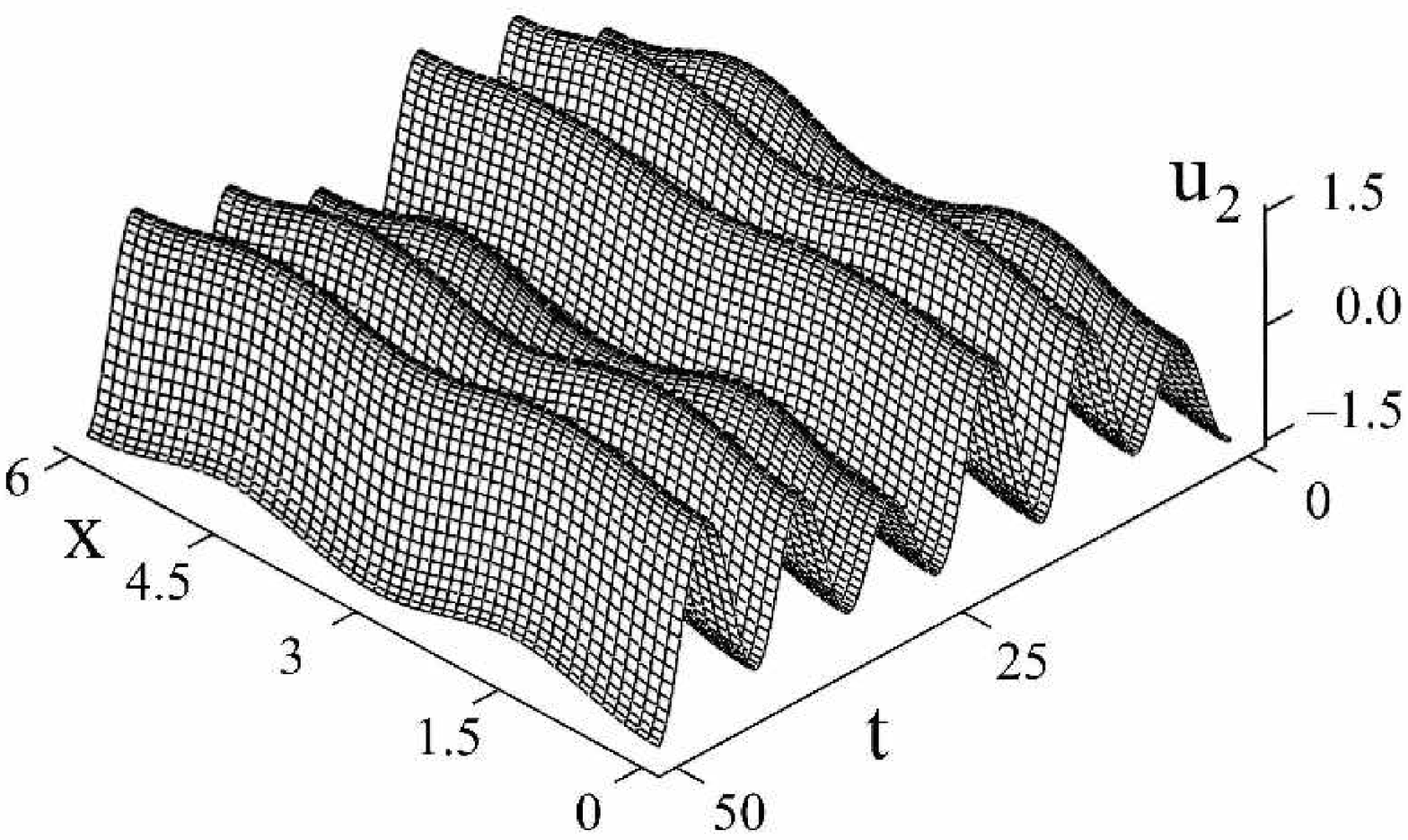}
\\
(a) & (b) \\
    &     \\
\includegraphics[width=0.25\textwidth]{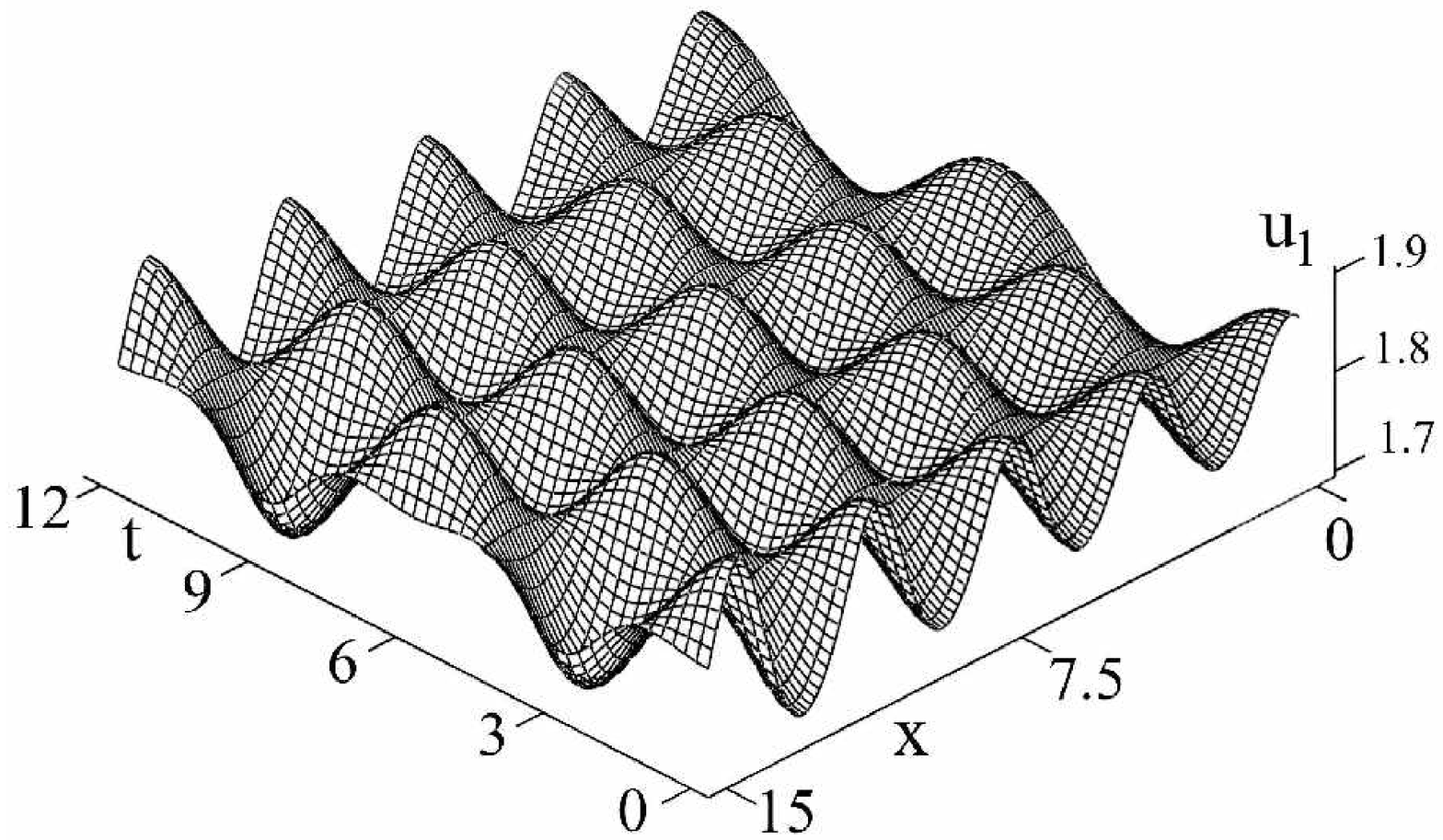}&
\includegraphics[width=0.25\textwidth]{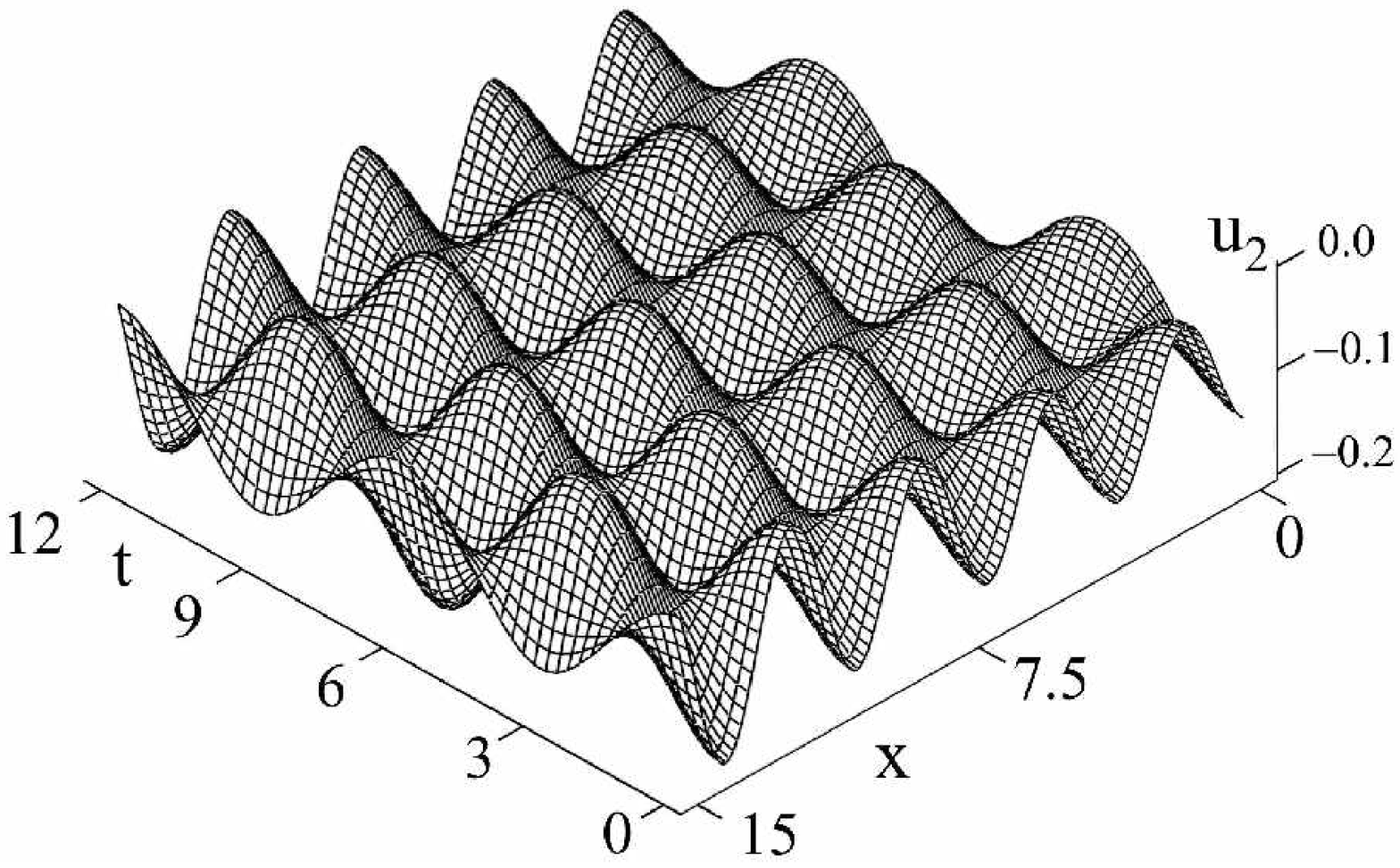}
\\
(c) & (d) \\
    &     \\
\includegraphics[width=0.25\textwidth]{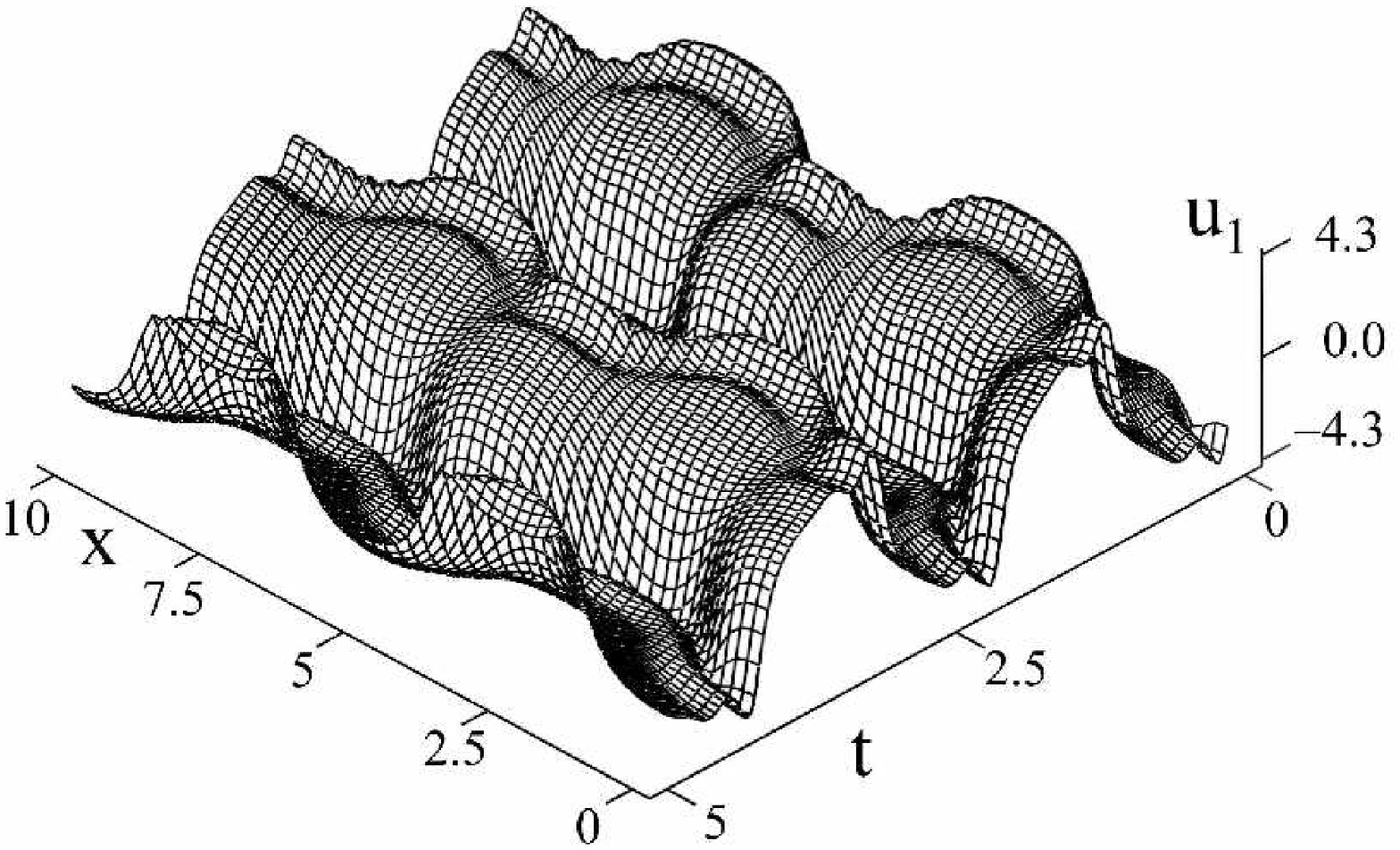}
&
\includegraphics[width=0.25\textwidth]{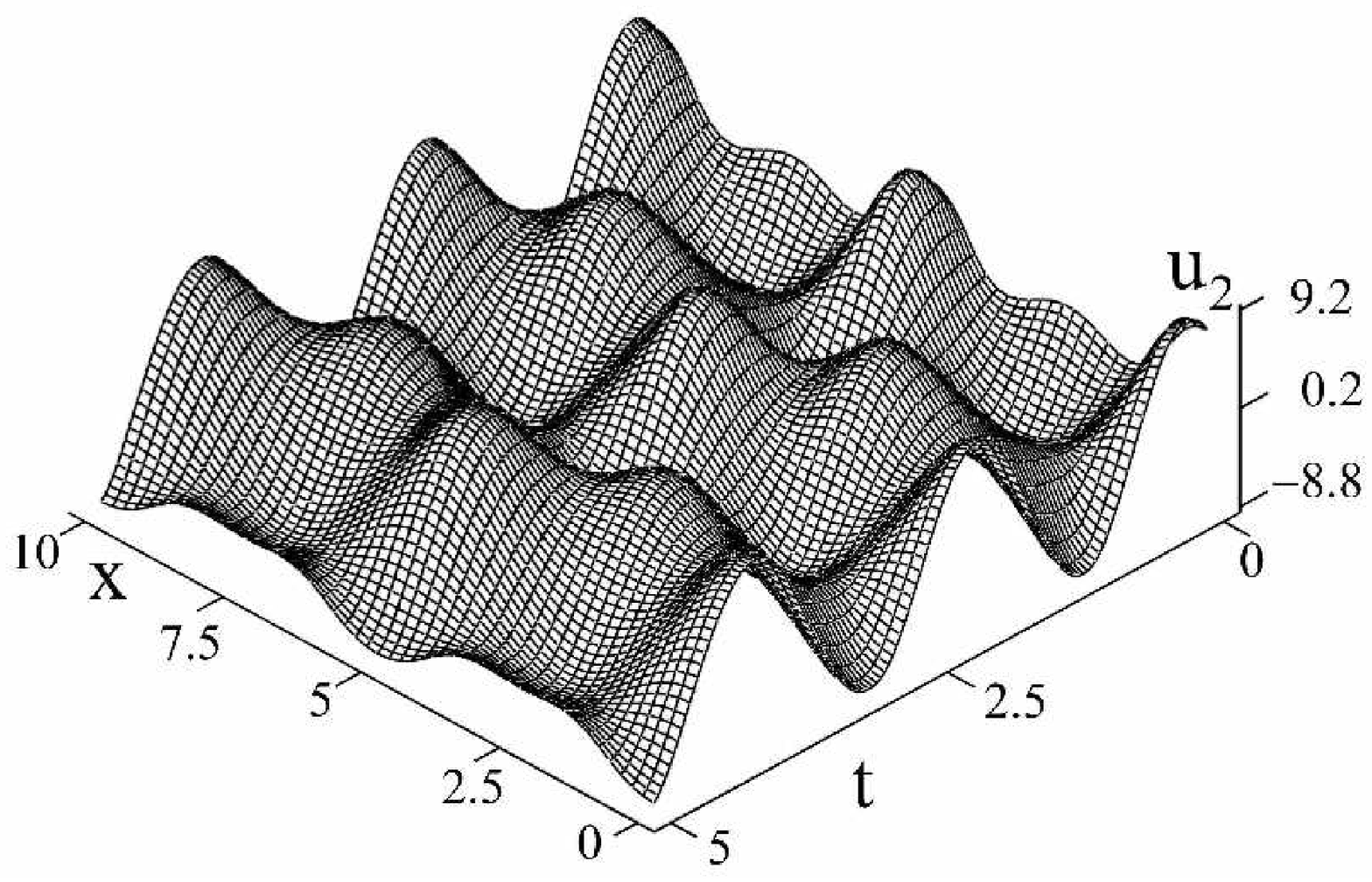}

\\
(e) & (f) \\
    &     \\
\end{tabular}%
\end{center}
\caption{Dynamics of pattern formation for $u_{1}$ (left column)
and $u_{2}$ (right column) variables. The results of computer
simulations of the systems at parameters: $A=-0.01$, $\protect\alpha =1.8$, $\protect\beta =1.01$, $l_{1}=0.02$, $%
l_{2}=1$, $\protect\tau _{1}/\protect\tau _{2}=3.5$ -- (a-b); $A=1.95$, $\protect\alpha =1.82$, $\protect\beta =1.01$, $%
l_{1}=0.1$, $l_{2}=1$, $\protect\tau _{1}/\protect\tau _{2}=0.6$ -- (c-d); $A=-0.01$, $%
\protect\alpha =1.75$, $\protect\beta =10$, $l_{1}=0.05$, $l_{2}=1$, $%
\protect\tau _{1}/\protect\tau _{2}=0.05$ -- (e-f); }
\label{rys11}
\end{figure}

Let us consider the bifurcation diagram presented in Fig. 1b,d. It
was
already noted that the region inside the curve is unstable for wave numbers $%
k=0$ and outside - it is stable. From the viewpoint of homogeneous
oscillations, the system is stable near $\overline{u}_{1}=0$.
However, if we have $l_{1}<<l_{2}$, the system becomes unstable
according to Turing instability. As a result, we expect the
formation of stationary inhomogeneous structures. In fact, at the
beginning, only inhomogeneous fluctuations grow in amplitude and
lead to inhomogeneous pattern formation. At the same time, at the
dynamics of structure formation, the amplitude of the structures
increases, and at maximum and minimum amplitude, the structures
fall into the domain where the homogenous structures are unstable.
As result, we obtain complex interaction of Turing and Hopf
bifurcations (Fig. 5a,b).

Inhomogeneous oscillatory structures are presented on pictures
(Fig. 5 c,d)). Due to different evolution pattern for two
variables, the activator variable $u_{1}$ is presented in the left
column and the inhibitor one $u_{2} $\ is presented on the right
column. Such structures are obtained at eigenvalues with negative
real part when the intersection of the null-cline
for activator variable is located on the decreasing part of null cline $%
u_{2}(u_{1})$.  The corresponding eigenvalues for this situation
are presented in Fig. 2b(v). From the eigenvalue plots (Fig. 2) we
can see
corresponding separated domains where inhomogeneous oscillatory modes with $%
k=2$ are unstable. Successive increase of the fractional
derivative index will increase the amplitude of the presented in
Fig. 5a,b  inhomogeneous oscillation. As a result, ingomogeneuos
oscillatory structures of large amplitude are realized in the
system.

Another type of oscillatory structures with a little bit more
complicated dynamics are presented in Fig. 5e,f for $\beta =10,$
$\tau _{1}<<\tau _{2}$ and $\alpha \lessapprox 2$. As in the case
considered above we have inhomogeneous oscillatory structures the
surface of which in the region of slow motion oscillates with fast
frequency $1/\tau _{1}.$\ Such behaviors are due to oscillatory
property of the activator system at $\alpha $ approaching the
value of $2$.

\section{CONCLUSION}

We have shown a complex spatio-temporal pattern formation in
simple FRD system and compared these results with standard one.
The fractional derivative index plays a crucial role in this
pattern formation because conditions of time bifurcations depend
substantially on its value. By
eigenvalue analysis we have studied instability conditions for $k=0$, and $%
k\neq 0$\ for different values of external parameter $\mathcal{A}$
(the same as $\overline{u}_{1}).$ Nonlinear solutions show that
dynamics of the system is determined by most unstable modes. When
linear increments are comparable, we have an interplay between
Hopf and Turing modes leading to more complex dynamics.

When a fractional derivative index is changed from $0$ to $1$, the
large-amplitude structures are stationary if a limit cycle is damped at $%
\tau _{1}\simeq \tau _{2}$. Oscillatory structures at these values
of fractional derivative index can be realized only at $\tau
_{1}<<\tau _{2}.$

When a fractional derivative index is changed from $1$ to $2$, the
large-amplitude structures have more complex dynamics. Moreover,
the spatiotemporal structures are observed even at $\tau
_{1}>>\tau _{2}$.\ Complex structures are observed in the region,
when the bifurcation parameter leads to Turing and Hopf
instabilities, as well as in the regions where these instabilities
are damped. The system moves to large amplitude limit cycle with
further change in the fractional derivative index.

\bibliographystyle{plain}
\bibliography{asme2e}

\begin{thebibliography}{99}
\bibitem{pr} Nicolis, G., Prigogine, I., 1997, \textit{Self-organization in
Non-equilibrium Systems}, Wiley, New York.

\bibitem{ch} Cross, M. C. and Hohenberg, P. C., 1993, ''Pattern formation
outside of equilibrium'', \textit{Rev. Mod. Phys.}, Vol.
\textbf{65}, pp. 851-1112.

\bibitem{KO} Kerner, B.S., Osipov, V.V., 1994, \textit{Autosolitons},
Kluwer, Dordrecht.

\bibitem{hw1} Henry, B.I., Langlands, T.A.M. and Wearne, S.L., 2005,
''Turing pattern formation in fractional activator-inhibitor
systems'', \textit{Phys. Rev. E}, Vol. \textbf{72}, 026101(14 p.)

\bibitem{lhw} Langlands, T.A.M., Henry, B.I. and Wearne, S.L., 2007,
''Turing pattern formation with fractional diffusion and
fractional reactions'', \textit{\ J. Phys.: Condens. Matter}, Vol.
\textbf{19}, 065115 (20 p.)

\bibitem{hw2} Hernandez, D.,  Varea, C. and  Barrio, R. A., 2009,
''Dynamics of reaction-diffusion systems in a subdiffusive
regime'', \textit{\ Phys. Rev. E}, Vol. \textbf{79}, 026109 (10
p.)

\bibitem{gd} Gafiychuk, V., Datsko, B., 2006, ''Pattern formation in a
fractional reaction-diffusion system'', \textit{Physica A}, Vol.\textbf{365}%
, 300-306.

\bibitem{gd07} Gafiychuk, V., Datsko, B., 2007, ''Stability analysis and
oscillatory structures in time-fractional reaction-diffusion
systems'', \textit{Phys. Rev. E }, Vol. \textbf{75}, 055201(4
p.)(R)

\bibitem{gd08} Gafiychuk, V. and Datsko, B., 2008, ''Inhomogeneous
oscillatory structures in fractional reaction-diffusion systems'', \textit{%
Physics Letters A}, Vol. \textbf{372}, pp.619-622.

\bibitem{cam} Gafiychuk, V., Datsko,B., Meleshko,V., 2008,
\textquotedblright Mathematical modeling of time fractional
reaction-diffusion systems\textquotedblright , \textit{J. Comp. Appl. Math.}%
, Vol. \textbf{220}, pp. 215-225.

\bibitem{siam} Golovin,A. A., Matkovsky, B. J., Volpert, V. A., 2008,
''Turing pattern formation in the brusselator model with
superdiffusion'', \textit{SIAM J. Appl. Math.}, Vol. \textbf{60},
pp. 251-272.


\bibitem{Nec} Nec, Y., Nepomnyashchy, A. A. and Golovin, A. A., 2008,
''Oscillatory instability in super-diffusive reaction-diffusion
systems:
Fractional amplitude and phase diffusion equations'', \textit{EPL }%
, Vol. \textbf{82}, 58003 (6 p.).

\bibitem{zz} Zaslavsky, G.M., 2002, ''Chaos, fractional kinetics, and
anomalous transport'', \textit{Phys. Rep.}, Vol. \textbf{371}, pp.
461-580.

\bibitem{kl} Metzler, R. and Klafter, J., 2000, ''The random walk?s guide to
anomalous diffusion: a fractional dynamics approach'',
\textit{Phys. Rep.}, Vol. \textbf{339}, pp. 1-77.

\bibitem{mach} Agrawal, O. P., Tenreiro Machado, J. A., Sabatier, J., 2007,
\textit{Advances in Fractional Calculus : Theoretical Developments
and Applications in Physics and Engineering,} Elsevier.

\bibitem{KST} Kilbas, A. A., Srivastava, H. M., Trujillo, J. J., 2006,
\textit{Theory and Applications of Fractional Differential
Equations,} Elsevier.

\bibitem{Uch} Uchaikin, V. V., 2008, \textit{Fractional derivative method,}
Artishok (in Russian).

\bibitem{ip} Podlubny, I., 1999, \textit{Fractional Differential Equations},
Academic Press.

\bibitem{skm} Samko, S.G., Kilbas, A. A. and Marichev, O. I., 1993, \textit{%
Fractional Integrals and Derivatives: Theory and Applications},
Gordon and Breach, Newark, N.J.

\bibitem{fh} R. FitzHugh, Biological Engineering, McGraw-Hill (1969), 1-85.


\end{thebibliography}

\end{document}